\documentclass[a4paper,12pt]{article}
\usepackage{graphicx}
\begin{document}

\title{MEASUREMENTS OF THE GRAVITY WAVES VELOCITY}
\author{V.A.Dubrovskiy}
\date{\textit{Institute of Geosperes Dynamics, Russian Academy of Science,
38 k.6 Leninskiy pr., Moscow 117334, Russia}}
\maketitle

\begin{abstract}
Some results are presented of the Earth's microseismic background.
It is assumed that background peaks should correspond to the resonance
gravity-wave exchange in the system of two gravity-connected
bodies.  The microseismic spectrum is compared with the distribution of
gravity potential of the nearest stars.  A close peak-to-peak correspondence
is found.  This correspondence and resonance condition lead to an evaluation
of the gravity-wave velocity.  The resulting value is nine orders of magnitude
more than the velocity of light.  Some consequences of such result are discussed.
\end{abstract}

\section*{Introduction}
The observational data presented of the Earth's microseismic background
associated with gravitational waves shows an influence of the nearest cosmic
objects on the Earth.  I will use the simplest theoretical model to avoid
complexities that might arise in a more sophisticated approach.  Let us focus
on two gravity-connected objects such as Earth-Moon, Earth-Jupiter, Earth-Saturn,
Earth-Sun, Earth-nearest stars.  These coupled pairs will be considered as antenna
and the whole earth as receiver or sensor.  Now if the gravity wave length is
comparable with the antenna dimension (the distance $L$ between the gravity-connected
objects), resonance phenomena may appear in the antenna, producing a peak in
the microseismic background spectrum.  

The source of the gravity waves with different frequencies in Univers may be the 
gravity waves exchange between gravity-connected objects and gravity instability of 
the cosmic clouds that leads to the stars formation. It is possible to show as an 
example mechanism of the gravity waves generation with broad frequency spectrum. 
There is a theorem: if some system can have the position of the unstable equilibrium 
between stable state and unstable one then this system can oscilate in stable area 
with low frequencies and this frequencies decreases when the system approaches to 
the equilibrium (threshold of the instability) with wavenumber being finite at zero 
frequency~\cite{ref1, ref2}. This theorem is aplicable 
exactly to the case of the gravity instability clouds in Universe. This well known 
Jeans's instability lead usually to the process of the stars formation~\cite{ref3}. 
Presented process must be preceded by the intensive gravity waves creation and 
spectrum of this waves will continuosly shifts in low frequancy range when clouds will 
aproach to the instability threshold. Consequently we can state that the gravity 
waves with different frequencies are existing in Univers allways as the stars 
creation process is existing allways.

It is possible to evaluate the gravity wave velocity $C$ using the simplest relation 
between frequency $\nu$ and wavelength $\lambda$, i. e., $\nu=C/\lambda$ (assuming 
$\lambda/2\sim L$) if such peaks are observed. Consequently, the goal of this study 
of the microseismic background is to search for such peaks.

\section*{Observations}  
Our observations were made in the seismic station at Simpheropol University at 
Sevastopol using laser interferometery~\cite{ref4}.  Six peaks were found at 
$2.3 Hz$, $1 Hz$, $0.9 Hz$, $0.6 Hz$, $0.4 Hz$, $0.2 Hz$ (see Fig.1~$a$,~$b$).  
These figures were taken directly from the spectranalyser SK4-72 using recording 
equipment. The SK4-72 accumulates the output interferometer signals and
enhances the periodic components of the signal relative to chaotic
components.  One thousand twenty four records of length 40 seconds
were averaged. 

There exist massive gravity objects at distances of 1.3, 2.7, 3.5, 5, 8, and around 
11 parsecs.  All the distances $L$ between earth and these objects correspond to all 
the observed peaks only if the value of the gravity-wave velocity $C$ is nearly nine 
orders of magnitude greater than the velocity of light, if we ascribe
\begin{figure}
\centering
\includegraphics[width=0.70\textwidth]{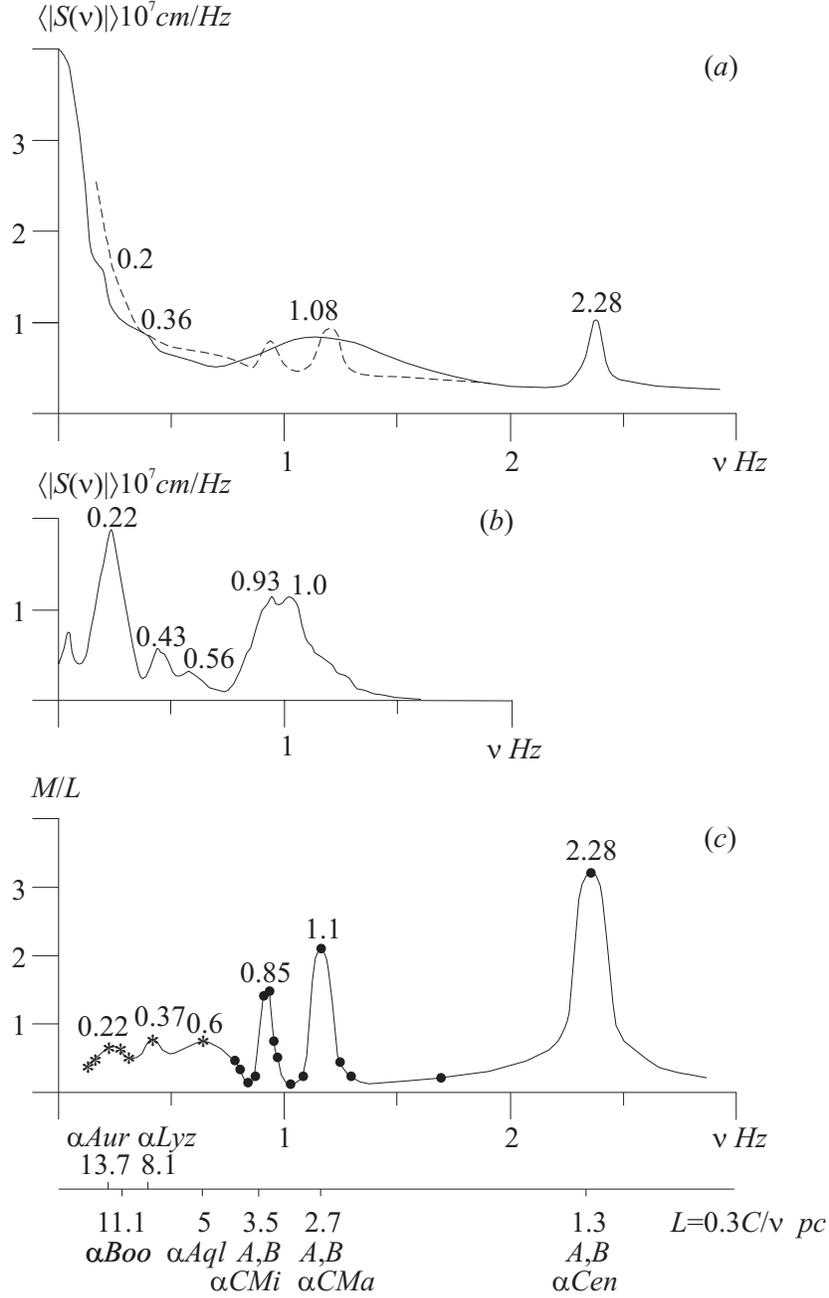}
\caption{Observed microseismic background (solid curve) after accumulation of the 
background signals from the interferometer output: $a$) - the $0.1-5 Hz$ range; $b$) 
the $0.1-2 Hz$ range. The dotted curve on a) shows the calculated distribution of the 
stars gravity potential plus the uniform part of the microseismic background. This 
dotted curve is normalized so as to equalize the value of the dotted and solid curves 
at $2.28 Hz$.  $c$) - the calculated distribution of the stars gravity potential. The 
solid points correspond to all nearest stars with distances $L<4$ parsecs, and the stars 
to all brightest stars for which $L>4$ parsecs. Mass $M$ is in sun masses.  
$\alpha Aur - \alpha Cen$ means $\alpha$ stars of the constellation in standard 
astronomical notation~\cite{ref5, ref6}. $A$, $B$ signifies star doublets.
The numbers at the extremal curve points are frequencies.}
\end{figure}
them to resonances in the earth-object system in accord with the simple
formula $\nu=C/\lambda$ ($\lambda/2\sim L$ due to resonance). This conclusion permits 
the calculation of the gravity-potential distribution of the nearest stars.  This 
distribution is shown in Fig.1~$c$.  Comparison of Fig.1~$a$,~$b$ with Fig.1~$c$ 
reveal a close similarity of corresponding curves: each peak of Fig.1~$a$,~$b$ 
corresponds to a peak on Fig.1~$c$ and vice versa.  However, there are non-important 
differences that should be pointed out for clarity.  For distances $L>4$ parsecs, 
data were taken only for the brightest star, and the curve of gravity potential 
corresponding to this distance is lower than for the $L<4$ level on the theoretical 
Fig.1~$c$. Another difference consists of the presence of uniform growth for the 
low-frequency background component on experimental Fig.1~$a$,~$b$ which does not 
appear on Fig.1~$c$.  Usually this uniform component of the microseismic background 
is described by the law $A_{\omega}\sim 1/\omega^2$~\cite{ref7, ref8}. 

Besides the quantitative correlation of frequency peaks with the distribution of 
the nearest stars, there is qualitative correspondence as well.  Namely, the 
sharpest peak at $2.28 Hz$ corresponds mainly to the distance between earth and 
the nearest stars doublet $A$ and $B$  ($\alpha$ Centaurus~\cite{ref5, ref6}). 
The broader peak at $1 Hz$ on Fig.1 $a$, $b$ corresponds to the distances to 
stars distributed over the range from 2.4 to 3.8 parsecs~\cite{ref5, ref6}. 
The spectranalyzer SK4-72 averages all resonance peaks for
this $2.4-3.8$ range into one broad peak around $1 Hz$ (Fig.1~$a$). But the broad 
peak on Fig.1 $a$, under more careful study, splits into two peaks (Fig.1~$b$) if 
spectranalyzer SK4-72 is processing in the frequency range from $0.1$ to $2 Hz$ 
(the exaggeration of the frequency scale). This subdivision of the frequency range 
corresponds to the divisions of the group of stars from 2.4 to 3.8 parsecs into two 
subgroups situated around 2.7 and 3.5 parsecs (Fig.1~$c$). 

The gravity potential distribution of these subgroups plus the uniform background 
spectrum is shown on Fig.1~$a$ by the dotted curve.  Thus we see both quantitative 
and qualitative correlation of the frequency spectra of the microseismic background 
that can be associated with resonance gravity-wave exchange. It is these correlations 
that provide the possibility of evaluating the gravity-wave velocity which turns out 
to be nine orders of magnitude greater than the velocity of light.

 If the experimental results are considered to be meaningful then it is possible
to propose further more decisive observations.  Namely, it is reasonable to look
for resonance peaks corresponding to the gravitational wave exchange of the Earth
with the Moon ($\sim 240 MHz$), the Sun ($\sim 0.6 MHz$), Venus ($\sim 0.3-2.2 MHz$), 
Jupiter ($\sim 100-150 kHz$), and Saturn ($\sim 58-72 kHz$). Moreover, the peaks 
corresponding to Venus, Jupiter and Saturn should change their frequency in 
accordance with the changing distance between Earth and these planets during their 
orbital motion around the Sun.  Establishing such a correlation will be a crucial 
experiment, decisively supporting the results presented above concerning the enormous
gravity-wave velocity and the elastic model of the physical vacuum~\cite{ref9}.

\section*{Discussion}
Such enormous velocity gravity waves were discussed before from a different
physical perspective~\cite{ref10, ref11, ref12} 
and~\cite{ref9}.  Laplace gave a lower-limit evaluation of the gravity
propagation velocity using observational data on the stability of the Earth-Moon
system.  This lower limit was found to be $5\cdot10^8$ times greater than the 
velocity of light $c$~\cite{ref12}.  If we normalize the gravity 
velocity in Laplace's formula to that of the light velocity, then we get the 
unacceptable conclusion that the Earth-Moon system could exist only about 3000 years.
The gravity-wave velocity evaluation was given independently using theoretical
assumptions about the elastic model of the physical vacuum in~\cite{ref9}
also.  This evaluation of the velocities ratio is close to that of Laplace and is
approximately equal to $(e/mf^{1/2})^{1/2}$ where $f$, $e$, and $m$ are the 
gravitational constant, the electron charge and the proton mass~\cite{ref9}.  
I did not know of Laplace's result when~\cite{ref9} was written, and therefore 
a citation to Laplace's result is absent from that paper. 

Some problems appearing in connection with the huge gravity-wave velocity are 
discussed in~\cite{ref9} on the basis of the elastic model of the physical 
vacuum. The very large difference of the gravity and electromagnetic velocities 
means, first of all, that many electromagnetic and gravitational phenomena are 
practically independent and their mutual influence appears only through their 
appearance in some physical constants. This huge gravity-wave velocity also means 
that we are living in a practically static gravitational field.  Retardation effects 
become substantial only for distances comparable with the dimensions of the Universe, 
or for very large cosmic objects at least.  This point of view lends a better 
understanding to the problems of Universe structure and its evolution, as the 
interaction time between the components of the Universe is considerably less than 
its lifetime, i. e., components of the Universe are causally connected and we do not 
need to violate any thermodynamic laws. The existence of a velocity that is larger 
than the light velocity violates, of course, the relativistic invariance of special 
relativity theory.  But this violation causes only small changes in values of the 
physical constants (for example, $e$) only in the eighteenth decimal, as in the 
relativistic root $(1-(v/C)^2)^{1/2}$ appearing in the elastic model of the physical 
vacuum~\cite{ref9}, the ratio $v/C$ does not exceed $c/C\sim10^{-9}$ in the 
approach discussed.  In the elastic model, shear waves compare with electromagnetic 
waves, longitudinal waves with gravitational waves, and the particles are considered 
as singularities of the elastic body.  From the point of view of the singularity, 
the elastic body is emptiness.  Consequently the problem of their capture does not 
appear.  Every singularity has an eigenfrequency spectrum with complex 
frequencies~\cite{ref13, ref9, ref14}.  
The real part of the frequency is connected with the self energy (mass) and the 
imaginary part determines the lifetime.  Consequently, each spectrum mode has 
eigenfrequency and eigenattenuation, i.e., self energy and lifetime are 
interconnected.  So, we could in principle set up the mathematical problem of the 
elementary particle mass spectrum as a problem of the inhomogeneity self
oscillations the elastic body with the above-mentioned difference of shear and
longitudinal velocities having eigenfrequencies $10^9$ times more than 
eigenattenuation~\cite{ref13}.  Consequently, our model of the physical vacuum can
explain the slow decay of elementary particles~\cite{ref9} if this decay process 
includes radial self-oscillation of the cavity (model of vacancy).  This is one more 
consequence of the large longitudinal wave velocity of elastic ether and consequently
the large gravity wave velocity. The elastic model of the physical vacuum also predicts 
conversion of gravity waves into electromagnetic waves and vice-versa.  Again, the very 
large ratio of gravity to electromagnetic velocity prevents intensive energy transfer 
between gravitational and electromagnetic phenomena, as the coefficient of transformation 
is proportional to $\sim10^{-9}$.  Only for very large-scale phenomena can transformation 
effects become important: for example, only between gravity-connected large-scale objects
can there be intensive gravity-electromagnetic wave exchange that would lead to creation 
of a photon background. And this electromagnetic background could have footprints of
the resonance gravity-wave exchange similar to the microseismic background peaks
discussed above.  There are, of course, other effects associated with the large
difference between the gravity and electromagnetic wave velocities.  But these are
subjects of further study.  At the end, it should be noted that the existence of
a gravitational-wave velocity nine orders of magnitude greater than the electromagnetic
wave velocity may lead to vast theoretical and experimental consequences and to
a better understanding of nature.

\end{document}